\documentstyle[12pt]{article}

\newcommand{\newsection}[1]{
\vspace{10mm}
\pagebreak[3]
\addtocounter{section}{1}
\setcounter{equation}{0}
\setcounter{subsection}{0}
\setcounter{footnote}{0}
\noindent{\large\bf \thesection. #1}
\nopagebreak
\medskip
\nopagebreak}

\newcommand{\newsubsection}[1]{
\vspace{8mm}
\pagebreak[3]
\addtocounter{subsection}{1}
\noindent{ \bf \thesubsection\hspace{.1in} #1}
\nopagebreak
\vspace{2mm}
\nopagebreak}

\newcommand{\nc}{\newcommand}
\newcommand{\rc}{\renewcommand}

\nc{\bc}{\begin{center}}
\nc{\ec}{\end{center}}
\nc{\be}{\begin{equation}}
\nc{\ee}{\end{equation}}
\nc{\bea}{\begin{eqnarray}}
\nc{\eea}{\end{eqnarray}}
\nc{\bd}{\begin{displaymath}}
\nc{\ed}{\end{displaymath}}

\nc{\s}{\sum}
\nc{\wh}{\widehat}
\nc{\ra}{\rightarrow}
\nc{\la}{\leftarrow}
\nc{\rra}{\ \longrightarrow \ }
\nc{\lla}{\ \longleftarrow\ }
\nc{\llrra}{{\mathop{\longleftrightarrow}}}
\nc{\lra}{{\mathop{\leftrightarrow}}}

\nc{\vi}{Virasoro\ }
\nc{\dv}{$D$--Virasoro\ }
\nc{\dvr}{$D$-Vir\ }
\nc{\cdv}{$D$-Vir$_{\rm c}$}
\nc{\sdv}{$D$-Vir$^{(+,0)}$}
\nc{\csdv}{$D$-Vir$_{\rm c}^{(+,0)}$}
\nc{\al}{\alpha}

\nc{\til}[1]{\tilde{#1}}
\nc{\lb}[1]{\label{eqn:#1}}
\nc{\rf}[1]{(\ref{eqn:#1})}

\nc{\ie}{{\it i.e.}}
\nc{\cf}{{\it cf}}
\nc{\eg}{{\it e.g.\ }}

\rc{\l}{\langle}
\nc{\r}{\rangle}
\nc{\rvac}{|0\rangle}
\nc{\gvac}{g|0\rangle}
\nc{\lvac}{\langle 0| }
\nc{\Bl}{\Bigl\langle}
\nc{\Br}{\Bigr\rangle}

\nc{\bun}[2]{{#1 \over #2}}
\nc{\pa}{\partial}
\nc{\zp}{z\pa_z}
\nc{\pat}[1]{{\partial_{#1}}}
\nc{\ddx}[1]{{d \over d #1}}
\nc{\dxdy}[2]{{d #1 \over d #2}}
\nc{\ppx}[1]{{\partial \over \partial #1}}
\nc{\pxpy}[2]{{\partial #1 \over \partial #2}}
\nc{\dexdey}[2]{{\delta #1 \over \delta #2}}
\nc{\ppt}[1]{{\partial \over \partial t_{#1}}}
\nc{\pptt}[1]{{\partial \over \partial \tt_{#1}}}
\nc{\pxpt}[2]{{\partial #1 \over \partial t_{#2}}}

\nc{\comb}[2]{\left(
\begin{array}{c}
#1\\
#2
\end{array}
\right)
}

\nc{\rvec}[1]{\stackrel{\leftarrow}{#1}}
\nc{\stprod}[2]{{\stackrel{\leftarrow}{#1}\stackrel{\rightarrow}{#2}
-\stackrel{\leftarrow}{#2}\stackrel{\rightarrow}{#1}}}

\nc{\fw}{F_{\rm w}}
\nc{\aw}{A_{\rm w}}
\nc{\bw}{B_{\rm w}}
\nc{\cw}{C_{\rm w}}
\nc{\hw}{H_{\rm w}}
\nc{\abw}{(AB)_{\rm w}}

\nc{\xd}{{X^{\mbox{\tiny  D}}}}
\nc{\xdn}[1]{{X^{\mbox{\tiny  D}}_{#1}}}
\nc{\xs}{{X^{\mbox{\tiny  S}}}}
\nc{\xsn}[1]{{X^{\mbox{\tiny  S}}_{#1}}}
\nc{\omd}{{\Omega^{\mbox{\tiny  D}}}}
\nc{\omdn}[1]{{\Omega^{\mbox{\tiny  D}}_{#1}}}
\nc{\ld}{{L^{\mbox{\tiny  D}}}}
\nc{\ldn}[1]{{L^{\mbox{\tiny  D}}_{#1}}}
\nc{\cldn}[1]{{{\cal L}^{\mbox{\tiny  D}}_{#1}}}

\nc{\pbr}[2]{{\{ #1 , #2 \}}}
\nc{\mbr}[2]{{\{ #1 , #2 \}_{\mbox{{\tiny{\rm M}}}}}}
\nc{\sbr}[2]{{\{ #1 , #2 \}_{\mbox{{\tiny{\rm S}}}}}}

\nc{\lam}{{\lambda}}
\nc{\hf}{{1\over2}}
\nc{\qint}{{{q^{x}-q^{-x}}\over{q-q^{-1}}}}
\nc{\ze}{\zeta}
\nc{\tet}{\theta}
\nc{\ta}{$\tau$\ }
\nc{\w}{$w_{1+\infty}$\ }
\nc{\W}{$W_{1+\infty}$\ }
\nc{\kp}{KP-hierarchy}

\nc{\inv}{^{-1}}
\nc{\Tr}{{\rm Tr}\,}
\nc{\res}{{\rm res}\ }
\nc{\bac}{\begin{array}{c}}
\newcounter{tabnum}
\nc{\bi}{\bibitem}
\nc{\nn}{\nonumber}
\nc{\bkl}{B\"acklund}
\nc{\sch}{Schr\"odinger}
\rc{\thefootnote}{\fnsymbol{footnote}}

\nc{\kmprod}[2]{{\stackrel{\leftarrow}{#1}\stackrel{\rightarrow}{#2}}}

\begin{document}
\addtolength{\baselineskip}{.7mm}
\vspace*{.1cm}

\begin{flushright}
TMUP-HEL-9609 
\end{flushright}

\bc
{\large\bf Difference Operator Approach to the Moyal Quantization\\[1.5mm]
 and
 Its Application
to Integrable Systems}\footnote[1]{to be published in
J. Phys. Soc. Jpn.} \\[2.3cm]
{ Ryuji Kemmoku}\footnote[7]{E-mail:\ kemmoku@phys.metro-u.ac.jp}\\[3mm]
{\it Department of Physics,\
Tokyo Metropolitan University\\
Minami-Osawa, Hachioji, Tokyo 192-03, Japan}\\[60mm]

{\bf Abstract}
\ec
Inspired by the fact that the Moyal quantization is related with
nonlocal operation,
I define a difference analogue of vector
fields and rephrase quantum description on the phase space.
Applying this prescription to the theory of the KP-hierarchy,
I show that their integrability follows to the nature of their Wigner
distribution.
Furthermore the definition of the ``expectation value'' clarifies
the relation between our approach and
the Hamiltonian structure of the KP-hierarchy.
A trial of the explicit construction of the Moyal bracket structure
in the integrable system is also made.

\vskip 1cm

\newpage

\newsection{Introduction}

\vskip  .3cm

In the classical mechanics, dynamics can be described in
the geometrical context.
If we define a Hamiltonian vector field on the phase space (more generally
 the symplectic manifold), the time evolution of canonical variables
 can be recognized as an integral curve of the Hamiltonian vector field.
It is equivalent to a Lie derivative as an operation to a
function defined in the phase space.
In this sense the vector field generates a Lie algebra.

When we quantize some dynamical systems,
we usually consider the
Hilbert space instead of the phase space and replace the Poisson bracket
of observables by a commutator of operators on the Hilbert space.
This is nothing but Dirac's correspondence principle.

Meanwhile effort to realize quantization on the phase space
has been also made.
Weyl associated quantum mechanical operators to classical functions of
position and momentum \cite{weyl}.
Wigner introduced the quantum ``distribution function'' (Wigner distribution)
and formulated the quantum statistics on the phase space \cite{wig}.
These two concepts were combined by Moyal \cite{moy}.
He showed that if one constructs a classical function along Weyl's
correspondence, an expectation value of it, which is given by using the
Wigner distribution, can be identified with one of the corresponding
quantum mechanical operator.
On the basis of these works, many authors have contributed to the progress in
formulation of the phase space quantization
({\it e.g.} see \cite{kub}--\cite{om}).

One of remarkable things of Moyal's work is the introduction of the
so-called Moyal bracket.
It plays a role of the commutator in the phase space quantization
and reduces to the Poisson bracket in the classical limit.
In the ordinary procedure the correspondence between the
classical and quantum mechanics is somehow mysterious since there
exists no explanation how such a transition takes place in the nature.
But the phase space quantization seems to clarify such mechanism:
As Moyal suggested in his paper, the Moyal bracket seems to be constructed
by using some nonlocal operators.
On the other hand,
the uncertainty principle implies the existence of a minimal lattice size
of the phase space.
{}From these two facts we arrive at the speculation that some {\it difference}
operator plays a key role in the Moyal quantization.

In Section 2, along this spirit, I first consider the difference analogue of
the Hamiltonian vector field and represent the Moyal bracket by using it.
This new vector field has a tricky property.
The basis of it consists of the difference operator.
For this reason a constant vector plays the role of indices which
originally associate with the local coordinates.
Then the dimension of the new vector field becomes infinite
even if one of the phase space is finite as it will be seen.
Moreover for the purpose of reformulating the Moyal quantization
in our prescription,
I regard the difference analogue of the Hamiltonian vector field
as an operator on the phase space and define its dual form.
The dual pairing provides us the measurement of the quantum mechanics
in a form of expectation value in the phase space.
I also discuss dynamics through the time evolution of the expectation
value in a simple case.
All these reproduce the well-known results in ordinary quantum mechanics.

Although until now I mention only about the Moyal quantization,
we already know other examples in which the difference
version of a system plays an important role.
One of them is nonlinear integrable systems.
The existence of difference analogue of integrable systems is not trivial
at all, since in general a naive discretization of continuous variables
will not preserve integrability but creates chaos in an arbitrary
nonlinear system.
The transition between integrable and non-integrable discretizations is
subtle and difficult to clarify the mechanism.
It is, however, supposed that a large symmetry exists behind
the integrable discrete system. We know, for instance,
Hirota's bilinear difference
equation \cite{hir}, which is a difference version of the KP-hierarchy
\cite{miw}, exhibits large symmetry explicitly in an
equivalent form of the Pl\"ucker relation \cite{sat}.
For other types of integrable discretization, see, e.g., \cite{jac}.
Besides these well-known examples we have studied the role of the
difference version of the integrable system.
In \cite{jpa}, we clarified that the $W_{1+\infty}$ symmetry \cite{pop}
in the KP theory is deeply
connected with the difference analogue of a discrete version of the conformal
symmetry.
As mentioned above the difference analogue of Hamiltonian vector field has
infinite dimensionality.
This fact implies that we could apply such formulation to the integrable
systems.

On the basis of the investigation, in Section 3, a main part of
this manuscript, I try to apply the
difference operator approach stated in Section 2 to the KP theory.
There the phase space whose coordinates are regarded as
the spectral parameter is considered.
I first define a functional
 corresponding to the Wigner distribution
by using the Baker-Akhiezer function in the place of the quantum wave function,
and a quantity analogous to the
expectation value in the Moyal quantization.
We see that the ``distribution'' induces the KP solution
via the additional symmetric flow \cite{chen,orl}.
Furthermore the ``expectation value'' will suit the Hamiltonian
structure of the KP-hierarchy \cite{la}.
At the end of this section, the Moyal bracket structure emerges.

Section 4 is devoted to construct the Moyal bracket explicitly in the
integrable system.
To this end I first pay attention to a fact that the algebraic structure
of the new vector field is analogous to a discretization of
the conformal symmetry.
Applying it with the case of the KdV equation, I will provide
a new bracket through the Moyal-like deformation of the Poisson structure
of the KdV equation \cite{gd}.

In the last section we summarize the results of this manuscript and
provide future perspective.

Finally three appendices are given to supply details of our
discussions.

\newpage
\newsection{Difference operator approach to the Moyal quantization}
\vskip .3cm

In this section I explain the essence of the difference operator
approach to the Moyal quantization based on the work in \cite{jpsj}.

\newsubsection{Difference analogue of vector fields}

Let $M$ be a differentiable manifold (${\rm dim}M=m$).
We take a local coordinate system $\vec x=(x_1,\cdots,x_m)$
on $M$ and define a (generalized) difference operator as
\be
\nabla_{\vec a}:= {1\over \lambda}\sin (\lam\sum_ja_j\partial_{x_j}),
\lb{sabun}
\ee
where $\vec a$ denotes a constant vector and $\lam$ is a parameter.
In the $\lam\ra 0$ limit, \rf{sabun} becomes $\s a_j\pa_j$.
It is the ordinary vector field.
This fact leads to the natural definition of
a new vector field whose basis consists of $\nabla_{\vec a}$.
I consider the following form:
\be
\xd=\int d\vec a\ v_\lam (\vec x,\vec a)\nabla_{\vec a},
\lb{vector field}
\ee
where $v_\lam$ is the component of $\xd$ in the local coordinate
system $\vec x$
on $M$.
Comparing with the ordinary vector fields of the
differential geometry, we see that the constant vectors $\vec a$ play
the role of the indices $j$ of the local coordinates $x_j$.
In this sense, $\xd$ is infinite-dimensional.
{}From now we regard \rf{vector field} as a difference analogue of the
ordinary vector field.

Starting from \rf{vector field},
we can discuss the ``difference
geometry'' by defining other geometrical objects:

\noindent\underline{Difference one-form $\omd$}
\be
\omd:= \int d\vec a\ w_{\lam}(\vec x, \vec a)\,\Delta^{\vec a},
\ee
where $\Delta^{\vec a}$ is the conjugate of $\nabla_{\vec a}$ satisfying
$
\l \Delta^{\vec a'},\nabla_{\vec a}\r=\delta(\vec a'-\vec a).
$
The one-form $\Delta^{\vec a}$ is, for example, realized as the
pseudo-differential operator
\be
\lam\csc\{\lam(\vec a\cdot\vec\partial)\} := {2i\lam\over e^{i\lam\vec
a\cdot\vec\partial}-e^{-i\lam\vec a\cdot\vec\partial}}=
2i\lam\sum_{n=0}^\infty
e^{-i(2n+1)\lam\vec a\cdot\vec\partial}.
\lb{dual}
\ee
This enables us to define the bilinear pairing explicitly:
\bea
\l \Delta^{\vec a'},\nabla_{\vec a}\r:=
\lam\csc\{\lam(\vec a'\cdot\vec\partial)\}
\cdot \bun1\lam\sin\{\lam(\vec a\cdot\vec\partial)\}
\ \delta(\vec a'-\vec a)
=\delta(\vec a'-\vec a).
\eea

\noindent\underline{Difference two-form $\omdn 2$}
\be
\omdn 2=\int d\vec\al\int d\vec\beta\ w_{\lam}(\vec x: \vec\al,\vec\beta)\,
\Delta^{\vec\al}\wedge \Delta^{\vec\beta}.
\lb{2form}
\ee
Here $\wedge$ is nothing but the ordinary wedge product.
The forms of higher degree can be constructed by using the operation
$\Delta$ repeatedly.

\noindent\underline{Exterior difference operator $\Delta$}
\be
\Delta\omdn 2=\int d\vec\gamma\int d\vec\al\int d\vec\beta\
\nabla_{\vec\gamma}\,
w_{\lam}(\vec x:\vec\al,\vec\beta)\,\Delta^{\vec\gamma}\wedge\Delta^{\vec\al}
\wedge\Delta^{\vec\beta}.
\ee
Since $[\nabla_{\vec\al}, \nabla_{\vec\beta}]=0$, $\Delta$ has the
desired property $\Delta\Delta=0$.

\noindent\underline{Interior product $i_{\nabla_{\vec a}}$}
\be
i_{\nabla_{\vec\gamma}}(\Delta^{\vec\al}\wedge\Delta^{\vec\beta})
=\delta(\vec\gamma -\vec\al)\Delta^{\vec\beta}
-\delta(\vec\gamma -\vec\beta)\Delta^{\vec\al}.
\ee

\noindent\underline{Lie {\it difference} ${\cal L}_\xd$}
\be
{\cal L}_\xd=\Delta\cdot i_{\xd} +i_{\xd}\cdot\Delta.
\ee

We obtain relations among operators $\Delta, i_{\xd}$, and ${\cal L}_\xd$
in the similar form to the continuous case.
This means that the operators $(d,i_X,{\cal L}_X)$ in the
differential geometry
are replaced by $(\Delta, i_{\xd}, {\cal L}_\xd)$ here.

\newsubsection{Difference operator approach to the Moyal Quantization}

Now let us consider the case that $M$ is symplectic, {\it} i.e.
the phase space.
I only treat the two-dimensional physical phase space $\vec x=(p,x)$
for simplicity. The following discussion is easily generalized to the higher
dimensional case.

In the classical mechanics, the Poisson bracket is geometrically
realized as the action of the Hamiltonian vector field on a function
such as $X_f\,g=-\pbr fg$.
Then it seems natural to think that an analogous formulation
for the Moyal bracket should exist.
I will show that the difference analogue of vector fields enables us to
realize such concept.
To this end
I first propose a difference version of the Hamiltonian vector field
 $X_f$ as follows:
\be
\xdn f=\int d\vec a\ v_\lam[f](\vec x, \vec a)\nabla_{\vec a},
\lb{dif}
\ee
where
\be
v_\lam[f](\vec x,\vec a)={1\over (2\pi)^2}\int d\vec b\
e^{i(\vec a\times\vec b)}f(\vec x +i\vec b).
\ee
In this expression $\vec a\times\vec b$ means the vector product of
$\vec a$ and $\vec b$, and it
inherits the symplectic structure
from the phase space.
If we apply $\xdn f$ on a function $g$, we have the following desired
 expression:
\be
\xdn fg=-\bun 1{\lam}
    \left.  \sin \biggl[\lam
    \biggl(\ppx {x_1}\ppx{p_2}-\ppx {p_1}\ppx{x_2} \biggr) \biggr]
      f(p_{1},x_{1}) g(p_{2},x_{2}) \right|_{p,x}=:- \mbr fg\ .
\lb{Moyal}
\ee
Therefore $\xdn fg$ can be regarded as a
difference operator representation of the Moyal bracket.
In the $\lam \ra 0$ limit, \rf{dif} becomes $X_f$
and \rf{Moyal} the Poisson bracket.
Moreover the relation as
\be
[\xdn f, \xdn g]=-\xdn {\mbr fg}
\lb{comm}
\ee
is verified
by explicitly using \rf{dif}.
In this sense $\xdn f$ is meaningful as the vector field on the
phase space.
This form of algebra has been studied in other context \cite{ffz} and
is sometimes called the Moyal bracket algebra
(or the sine algebra).

Remark that the difference two-form corresponding to
the symplectic form $\omega$ can be considered:
\be
\Omega:=\bun 1{2i\lam}\int da\int db\ e^{-i\lam(\vec a\times\vec b)}\,
\Delta^{\vec a}\wedge\Delta^{\vec b}.
\lb{dsymp}
\ee
For the two-form we can show
\be
i_{\xdn f}\Omega=\Delta\,f.
\ee
This relation is analogous to the definition of the Hamiltonian vector field
as $i_{X_f}\omega=df$.
Therefore I interpret \rf{dsymp}
as a difference version of the symplectic form.

Next I establish the way to reconstruct the Moyal quantization method
by identifying the parameter
$\lambda$ with Planck's constant as $\lambda=\hbar/2$.
There are several methods which connect an (quantum) observable on the
Hilbert space with a function on the phase space.
If we introduce the Moyal bracket as a quantum version of the Poisson bracket,
we must fix the ordering of the operators along the so-called Weyl
correspondence \cite{weyl}.
On the basis of the procedure of the phase space quantization, I regard
$\xdn A$ as the object corresponding to the phase space function $A$.
This means that $\xdn A$ takes the place of the observable in our description.

In addition to this, an expectation
value of $\xdn A$ must be introduced and it should be equivalent to
the expectation value
in the ordinary quantum mechanics.
For this purpose I define the one-form associated with the
Wigner distribution $\fw$ \cite{wig} (see Appendix A) by
\be
\Omega_{\fw}=\int d\vec a\int d\vec b\ e^{-i(\vec a\times\vec b)}
\fw(\vec x+i\vec b)\,\Delta^{\vec a}.
\ee
Using the one-form and the orthogonality between $\nabla_{\vec a}$
and $\Delta^{\vec a}$, we actually obtain
\be
\l \Omega_{\fw}, \xdn A\r
=\int d\vec x\ \fw(\vec x)\,A(\vec x).
\lb{expect}
\ee
The right-hand side is nothing but the expectation value of $A$ in the
phase space quantization.
Hence we can say that the left-hand side is a new description of
 the expectation value.
(See \cite{jpsj} for more detail.)

We can also consider the time evolution of \rf{expect}.
The time dependence of $\l \Omega_{\fw},\xdn A\r$ can be
two fold: in the Heisenberg picture we have $\l \Omega_{\fw},\xdn A\r_t=\l
\Omega_{\fw},\xdn A(t)\r$,
while in the Schr\"odinger picture we have
$\l \Omega_{\fw}(t),\xdn A\r$, which must be equivalent.
(In the following discussion, we assume that $\fw$ (in Heisenberg) and
$A$ (in Schr\"odinger) have no explicit time dependence.)
If we take the Heisenberg picture, the time evolution of a physical observable
\rf{moyeq} of Appendix A should follow to one of $\xdn A$.
In fact, if the static Hamiltonian $H$ is given, the equation
\be
{d\over dt}\xdn A=-[\xdn A, \xdn H]
\lb{equation of motion}
\ee
enables us to make such an interpretation.
Since $\xdn A(t)$ is written as
$\xdn{A(t)}$, this equation is physically equivalent to \rf{moyeq} in
Appendix A,
and $A(t)=e^{\xdn Ht}A$.
Then from \rf{comm}, we obtain
\be
{d\over dt}\l \Omega_{\fw},\xdn A(t)\r
=\l \Omega_{\fw},\xdn {\mbr{A(t)}H}\r.
\lb{timeevo}
\ee
The right-hand side of \rf{timeevo} is identical with
$\l \Omega_{\mbr H{\fw(t)}},\xdn A\r$  (see Appendix B).
Hence we obtain an equation which $\Omega_{\fw}(t)=\Omega_{\fw(t)}$
must satisfy in the
Schr\"odinger picture:
\be
\ddx t \Omega_{\fw}(t)=\Omega_{\mbr H{\fw(t)}}.
\ee
Here we used $\fw(t):=e^{-\xdn Ht}\fw$.
It is clear that this equation corresponds to \rf{wigeq}.

All these are our reconstruction of the Moyal quantization.
Though these results have been known by using other methods,
the difference operator approach must provide a new insight
into the Moyal quantization not only formally but also practically.

\newsection{Application to the \kp}
\vskip .3cm

In the difference analogue of the vector field approach, we notice that
it has property different from the ordinary vector field,
{\it i.e.}
infinite dimensionality.
Even if we start from the finite-dimensional phase space,
its quantum version leads us to an infinite-dimensional vector field
in a sense that the vector $\vec a$ plays the role of the indices.
Since $\{\xd\}$ constitutes an infinite-dimensional
Lie algebra,
it may have some connection with symmetry of nonlinear
integrable systems.
The purpose of this section is to clarify it by applying our methods
to the theory of the \kp.


First we consider the complex-valued phase space $(z,\zeta)$ instead of
$(p,x)$ and a
function on it is assumed to be expanded in the formal Laurent
series as
\bd
A(z,\ze)=\s_{m\in{\bf Z}}\s_{n\in{\bf Z}}  a_{mn}\,z^m\ze^n\ .
\ed
In this case $\xdn A$ becomes
\be
\xdn A= \bun1{\lam}\s_{m,n}\ a_{mn}\,z^{m}\ze^{n}
\sin\Bigl\{ \lam(n\pa_{\ln z}-m\pa_{\ln\ze})\Bigr\}
=:\s_{m,n} a_{mn}\,z^{m}\ze^{n}\nabla_{mn}\ .
\ee
The factor $\nabla_{mn}$ plays a role of $\nabla_{\vec a}$ in
\rf{sabun}.
More precisely this expression is equivalent to the form that
the integration is carried out in \rf{sabun}.
If we define the ``components'' $\xdn{mn}$ of $\xdn A$ by
$z^{m}\ze^{n}\nabla_{mn}$, it constitutes a basis of the
following Lie algebra:
\be
[\xdn{m_1n_1}, \xdn{m_2n_2}]={1\over\lambda} \sin\Bigl\{\lam(n_1m_2-n_2m_1)
\Bigr\}\ \xdn{m_1+m_2,\,n_1+n_2}
\lb{zach}
\ee
\cite{ffz}, and also
that $\xdn{mn}$ satisfy
\rf{comm}\footnote{The Moyal bracket is now expressed as
\bd
\mbr fg= -\bun1\lam
\left.\sin\,\left\{\lam\biggl(\ppx {\ln z_1}\,\ppx {\ln {\ze}_2} -
\ppx {\ln {\ze}_1}\,\ppx {\ln z_2}\biggr)\right\}
\,f(z_1,{\ze}_{1})\,g(z_2,{\ze}_{2})\right|_{z,\ze}.
\ed}.

Next I define the corresponding object of the Wigner distribution.
In the theory of the \kp, the Baker-Akhiezer function
plays the role of the wave function of quantum physics
in the sense that it satisfies the linear equations of the inverse problem
associated with the nonlinear differential equations.
Then it is preferable to compose the ``distribution'' from the Baker-Akhiezer
function by regarding the spectral parameter as the coordinate of
the phase space $(z,\ze)$.
But we must remember that the KP-hierarchy is the system which has multi-time
evolution; and such evolution is generated by the operation of the
pseudo-differential operator.
Moreover the complex conjugate of the wave function in the quantum mechanics
is turned into the formal adjoint in this case.
Taking these facts into consideration, the following {\it functional}
 is suitable
for our purpose:
\be
F_{\rm KP}(z,\ze):=
\int dx\s_{l\in{\bf Z}}   w(q^{\bun l2} z)\,
w^{*}(q^{-\bun l2}z)\,\ze^{-l}
\lb{zwig}
\ee
where the parameter $\lam$ is fixed to $-i\ln q\ (|q|<1)$, and
$w\,(w^*)$ stands for the (adjoint) Baker-Akhiezer function.
The integration over the variable $y$ in \rf{wigd} of Appendix A
corresponds to the summation over\, $l$.
The integration over $x$ does not exist in \rf{wigd}.
This is because the Baker-Akhiezer function has $x$-dependence
(see \rf{ba} in Appendix C),
while in the quantum wave function there is no such dependence.
(Remark that the variable $x$ in the Baker-Akhiezer function
does not mean position since
in this case $z$ and $\ze$ are the coordinates of the phase space.)
Hence I smear out the $x$-dependence of the Baker-Akhiezer function
by taking an ``average''
in the definition of $F_{\rm KP}(z,\ze)$.

We can define the one-form $\Omega_{F_{\rm KP}}$ associated
with \rf{zwig} in the same way as discussed in the previous section.
It is clear that \rf{zwig} is formally expanded as
$
F_{\rm KP}=\s_{m,n}f_{mn}\,z^m\ze^n
$.
Then first we introduce the dual basis $\Delta^{mn}$ of $\nabla_{mn}$
and define $\Omega_{F_{\rm KP}}$ by
\be
\Omega_{F_{\rm KP}}(z,\ze)=\s_{m,n}f_{mn}\,z^m\ze^n\,\Delta^{mn}\ .
\ee
The ``expectation value'' now becomes
\be
\l \Omega_{F_{\rm KP}},\xdn A\r:=-\oint\bun{dz}{2\pi iz}\,\oint\bun{d\ze}
{2\pi i\ze}\ F_{\rm KP}(z,\ze)\,A(z,\ze)
=\s_{m,n}f_{mn}\,a_{mn}
\lb{zexp}
\ee
on the basis of the orthogonality relation as
$\l \Delta^{mn},\nabla_{m'n'}\r =\delta_{mm'}\delta_{nn'}$.
Here each integral means to pick up the coefficient of
$z^{-1}$ ($\ze^{-1}$, resp) term.
In particular if we use $\xdn{mn}$, we obtain
\be
f_{mn}=\l \Omega_{F_{\rm KP}}, \xdn{mn}\r
=
-\int dx\oint \bun{dz}{2\pi iz}\ z^{m} w(q^{\bun n2 } z)\,
w^{*}(q^{-{\bun n2}}z).
\lb{zfou}
\ee

It is interesting to consider this result in connection with the
additional flow $\pa_{kl}$ discussed in \cite{chen,orl}.
The use of $\pa_{kl}$ enables us to
represent the right-hand side of \rf{zfou} in a simpler form as
\be
\int dx\ {\cal D}_{mn}\,\biggl(-\ppx x \ln\tau\biggr).
\lb{diso}
\ee
See Appendix C for more detail.
In this expression, $\tau$ is the tau function of the KP-hierarchy and
${\cal D}_{mn}$ is defined as
\be
{\cal D}_{mn}:=
q^{\bun{nm}2}\s_{j,l=0}^{\infty}c_{jl}\bun{(n\lam)^j}{j!}
\, \pa_{m+l-1,l}
\lb{change}
\ee
where $c_{jl}$ is
\bd
c_{jl}=\s_{\al=1}^{l}\bun{(-1)^{l-\al}\ \al^{j}}
{(l-\al)!\ \al!}
\hspace{.1in} (j,l\geq1),
\hspace{.2in} c_{l0}=c_{0l}=\delta_{l,0}\ .
\ed

The quantity $-\pa_x\ln\tau$ is nothing but the solution of the \kp.
This means that $\Omega_{F_{\rm KP}}$ induces the KP solution.
On the other hand, ${\cal D}_{mn}$ comes from $\xdn{mn}$.
In this sense the additional
symmetry has been given its alternative interpretation
 from the discretization point of view.
Strictly speaking, the subscrips $m$ and $n$ run over the whole integer,
while one of the subscripts of the additional flow $\pa_{kl}$
the positive integers (see Appendix C).
Then the correspondence between ${\cal D}_{mn}$ and $\pa_{kl}$
is not one-to-one.
Nevertheless
 the essential point is that ${\cal D}_{mn}$ can be represented
by using the additional flow.
If we want to let the correspondence be one-to-one, we must truncate
the area over which one of the subscripts of ${\cal D}_{mn}$ runs.
Such a situation can be realized, for instance,
by imposing some condition on $a_{mn}$ and/or $f_{mn}$.
The additional
symmetry is known as a kind of the so-called $W$-symmetry \cite{walg}.
Therefore above consideration leads us to think that the discretization
which preserves integrability is
closely related to the $W$-symmetry in general
since
such symmetry is believed to possess
the universal properties of various integrable systems.

If we define ${\cal D}_A :=\s_{m,n}a_{mn}{\cal D}_{mn}$, we at last
obtain the ``expectation value'' as the following functional:
\be
\l \Omega_{F_{\rm KP}}, \xdn A\r=\int dx\ {\cal D}_A
\,\biggl(-\ppx x\ln\tau\biggr)
=: \til A(t).
\lb{atil}
\ee
Let us investigate the meaning of the functional $\til A(t)$.
The (multi-) time evolution of $\til A(t)$
is obtained by taking differential with respect to $t_r$:
\be
\ppt r {\til A}
=
-\int dx\ {\cal D}_A \biggl(\ppt r \ppx x \ln\tau\biggr)
=:-\int dx\ {\cal D}_A J_r.\nn
\lb{jr}
\ee
Here $J_r=\pa_r\pa_x\ln\tau$ is the first integral of the
KP-hierarchy, i.e. $H_r=\int dx\, J_r$ is the Hamiltonian of
the KP-hierarchy \cite{la}.
Since the non-commutative flow ${\cal D}_A$ generates the independent
symmetry from the ordinary KP-flow $\pa_r$, this equation means that
the time evolution of the ``expectation value'' naturally yields
the Hamiltonian structure of the KP-hierarchy.
For example, if we take the derivative with respect to another time variable
$t_{r'}$, we obtain
\bd
\ppt {r'} \int dx\ {\cal D}_A J_r=0.
\ed
In this sense, the right-hand side of \rf{jr} is nothing but the Hamiltonian
of the KP-hierarchy induced by the non-commutative flow.
This fact is not surprising:
The form such as $w\cdot w^*$ is known as the residue
(with respect to the pseudo-differential operator)
of a ``resolvent'' of the Hamiltonian mapping
in the KP-hierarchy \cite{la}.
And turning back to our definition of $F_{\rm KP}$,
we find that $F_{\rm KP}$ essentially consists of such bilinear form
although it is constructed by analogy with the Wigner distribution.
Hence the emergence of the Hamiltonian is natural.

On the other hand, we can introduce the vector field associated with the
functional $\til A$.
Using it, the right-hand side of \rf{jr} is formally defined as
\be
-\int dx\ {\cal D}_A J_r=:{\cal D}_{\til A}\cdot H_r\,.
\ee
The action of ${\cal D}_{\til A}$ on the functional $H_r$ has the same
structure as \rf{Moyal}.
Therefore we can interpret that ${\cal D}_{\til A}\cdot H_r$
provides the Moyal bracket structure in the KP theory, i.e.
we can write down \rf{jr} as follows:
\be
\ppt r \til A =
\{{\til A},{H_r}\}_{\mbox{{\tiny{\rm M}}}}^{\mbox{\tiny{(KP)}}}.
\lb{kpmoy}
\ee

I have shown that the difference operator
approach to the Moyal quantization
can be applied to the KP theory.
It is remarkable that the integrability of the KP-hierarchy
($\tau$ function, Hamiltonian structure)
is naturally led from the particular forms of $F_{\rm KP}$ and $\til A$.
On the other hand the algebraic structure of $\xdn f$ discussed in Section
2 restricts the form of $\Omega_{F_{\rm KP}}$.
{}From these facts I like to emphasize that the Moyal bracket structure
implies the origin of
the integrability of the system.


\newsection{Deformation of the Poisson structure of the soliton system}
\vskip .3cm

In the preceding section I revealed the Moyal bracket
structure in the KP system by considering the time
evolution of the functional $\til A$.
Then I arrive at a speculation that in integrable systems
 such structure can be constructed manifestly.
I will examine in this section whether it is appropriate.

First I notice that from \rf{change} we can associate
$\xdn{mn}$ with the holomorphic vector field.
For the holomorphic function $f(z)=\s f_m z^{m+1}$, the
operator which generates the infinitesimal conformal transformation
is defined as
\be
L_f=f(z)\,\pa_z=\s_{m\in{\bf Z}}f_m\, z^{m+1}\pa_{z}=:
\s_{m\in{\bf Z}}f_m\, L_m.
\lb{holv}
\ee
This can be understood as the vector field in the following sense:
\be
[L_f, L_g]=L_{L_f\,g-L_g\,f}.
\ee
The component $L_m$ is the basis of the Virasoro algebra.
By analogy with the case of the Hamiltonian vector field, I provide the
difference analogue of \rf{holv} by
\be
\ldn f:=\s_{m\in{\bf Z}}\s_{n\in{\bf Z}_{\geq0}} q^{-\bun{n(m+2)}2}\,f_m
\,z^{m}q^{n(\zp+\bun m2)}
=:\s_{m,n}q^{-\bun{n(m+2)}2}\, f_{m} \ldn{mn}.
\lb{dho}
\ee
Remark that the subscript $n$ is truncated to the non-negative integer
(see the discussion above).
In this case $\ldn f$ also constitutes a Lie algebra:
\be
[\ldn f,\ldn g]=\ldn{\ldn fg-\ldn gf}\,.
\lb{dholv}
\ee

When the coefficient of the function $A(z,\ze)=\s a_{mn}z^m\ze^n$
is given as
$a_{mn}=q^{-n(m+2)/2}\,a_m$, the ``expectation value'' of $A$
is written as follows:
\be
\l \Omega_{F_{\rm KP}},\xdn A\r=
\int dx\ \left[\oint \bun{dz}{2\pi iz}\ \biggl(\ldn A V(z)\biggr)\,
V^*(z)\right]\,\biggl(-\ppx x\ln\tau\biggr).
\ee
{}From this fact, \rf{dholv} is recognized as a discretization
of the holomorphic vector field which corresponds to the Moyal deformation.
Schematically, the correspondence is illustrated as
\bea
X_f\,g\ \ ({\rm Poisson})&\llrra& L_f\,g-L_g\,f\nn\\
\xdn f\,g\ \ ({\rm Moyal})\hspace{.08in}&\llrra& \ldn f\,g-\ldn g\,f \nn
\eea
In \cite{jpa} we showed that such type of
$q$-difference operator realization of
the \w algebra can be understood
as a specific realization of the discretization of the conformal symmetry.
(Actually it was constructed by using $\ldn{mn}$ in \rf{dho}.)
Also in this sense above interpretation is naturally achieved.
The appearance of the holomorphic vector fields also implies that
my procedure of the expectation value has mathematically similar
background to
the polarization process
in the geometric quantization \cite{wood}.
Then the relation between the Moyal quantization and the geometric
quantization might be clarified from our point of view.

The above diagram for the holomorphic vector fields
indicates the existence of some new bracket corresponding
to the Moyal bracket.
In the following discussion,
I will show such evidence by considering the
KdV equation $u_t-6uu_x+u_{xxx}=0$ as a simple example.
For the solution $u(x,t)$, let us define the functional ${\cal L}_m$ as
\be
{\cal L}_m=\int dx\ x^{m+1}\,u.
\ee
It is known \cite{gerve}
that the following bracket constitutes the Virasoro algebra:
\bea
\{{\cal L}_{m},{\cal L}_{n}\}&=&\int dx\ x^{m+1}(u\pa +\pa u)\,x^{n+1}
\hspace{1cm}(\pa=\pa/\pa x)\nn\\
&=&(m-n)\,{\cal L}_{m+n}.
\lb{gdvir}
\eea
This bracket has the properties of the Poisson bracket and
provides a part of the second Hamiltonian structure of
the KdV equation\footnote{More precisely, the term as $\pa^3$
must be added to $(u\pa +\pa u)$ in \rf{gdvir}
 and this term is nothing but the origin of the center.
But in our naive discussion here, we temporary ignore it.}.
The fact can be contrasted to the case of the holomorphic vector field.
Now I propose a functional corresponding to the difference
analogue of the vector field $\ldn{mn}$ in \rf{holv} as
\be
\cldn{mn}=\int dx\ q^{-n}x^{m}\,u
\ee
under the definition of the new bracket:
\be
\{\cldn{mn}, \cldn{m'n'}\}_q
:=\int dx\ q^{-n}x^{m}
(q^{n' x\,\pa}\,u\,q^{n x\,\pa}-
q^{{-n'} x\,\pa}\,u\,q^{{-n} x\,\pa})\,q^{-n'}\,x^{m'}.
\lb{qgd}
\ee
If we take the partial integration
\bd
\int dx\ f(x)\,(q^{x\pa}g(x))=\int dx\ (q^{-x\pa}f(x))\,g(x),
\ed
the right-hand side of \rf{qgd} yields
\be
(q^{nm'-n'm}-q^{-nm'+n'm})\,\cldn{m+m'\,n+n'}\ .
\ee
After substitution of $-i\ln q$ for $\lam$
and some change of normalization,
\rf{qgd} becomes the same algebra as \rf{zach}.
In this sense the term
\be
q^{n' x\,\pa}\,u\,q^{n x\,\pa}-
q^{{-n'} x\,\pa}\,u\,q^{{-n} x\,\pa}
\ee
can be interpreted as a deformation (discretization) of
the term as $u\pa+\pa u$.
More rigorous treatment may lead us to the Moyal-like deformation
of the Hamiltonian structure of the KdV equation.
Furthermore
it is interesting to generalize
the above investigation to the integrable hierarchy.
Recently
the $q$-deformation
of the integrable hierarchy was discussed in
\cite{frenkel}.
There the $q$-difference version of the pseudo-differential
 operator plays an essential role.
Such a generalization may have deep connection with it.


\newsection{ Summary and Discussion}

\vskip .3cm

First I summarize the results of the manuscript.

In section 2, the difference analogue of the
vector fields $\xd$ and its dual form $\Omega$ were defined;
and their geometrical properties were discussed.
In the reformulation of the quantum mechanics,
I proposed the difference Hamiltonian vector field $\xdn A$ and
its dual form $\Omega_{\fw}$ which was
associated with the Wigner distribution.
The expectation value of an observable could be constructed
by pairing of $\xdn A$ with $\Omega_{\fw}$.
The time evolution of observable and the Wigner distribution was
rewritten in our prescription.
All these results were already known in the context of the phase space
quantization.
Although I paraphrased them again by using the difference operator
and reformulated them since such construction had the possibility to
provide new insight into the Moyal quantization.

In section 3, I showed that the preceding formulation was
applicable to the theory of the KP-hierarchy:
The ``distribution'' $F_{\rm KP}$ was defined by using the Baker-Akhiezer
function in this case.
The one-form $\Omega_{F_{\rm KP}}$ is connected with the KP solution;
and the deformed vector field $\xdn A$ induces
the non-commutative (or $w_{1+\infty}$) flow of the KP solutions.
The time evolution of the functional $\til A$ which corresponds to the
``expectation value'' yields the Hamiltonian structure of the KP-hierarchy.
After all it has been clarified that the integrability of the KP-hierarchy
followed to the nature of $\Omega_{F_{\rm KP}}$ and of the
Moyal bracket.

In addition to these formal study of the Moyal
quantization, in Section 4,
I discussed $\xdn A$ was related to the algebraic
structure which was characteristic of integrable hierarchies:
The component $\xdn{mn}$ is identical with a realization of
 the difference analogue of the
conformal algebra.
Moreover this fact led us to the Moyal-like deformation of the Hamiltonian
structure of the KdV equation such as $\{\cldn{mn}, \cldn{m'n'}\}_q$.
Therefore the difference analogues of the
integrable system provided by the Moyal structure should
preserve their integrability.

\vskip .55cm

The Moyal bracket structure in the integrable system was also
discussed in other context \cite{str}.
There the variables $x$ and $z$ are regarded as the canonical variable.
Originally, such description was made in the theory of
 the dispersionless KP-hierarchy \cite{takebe}.
In this case, the pseudo-differential operator plays a role of
the quantum operator, i.e. the following correspondence
is realized:
\bd
P(x,\pa_x)=\s_j p_j(x)\,\pa_x^j\longmapsto {\cal P}(x,z)=\s_j p_j(x)\,z^j.
\lb{kup}
\ed
This correspondence is equivalent to
take other type of operator-ordering, so-called
the standard ordering \cite{meh}-\cite{kup}.
Nevertheless I think that the difference operator approach discussed
above is still effective here.
I will report it elsewhere.

\vskip 1cm

\noindent{\large \bf Acknowledgements}
\vskip .2cm

I would like to thank Satoru Saito for many valuable discussions
and careful reading of this manuscript.

\vskip 1cm

\appendix

\noindent{\large\bf Appendix \hspace{-.4in}
\newsection{\hspace{.15in}}}

\vskip -.8cm

In this appendix I briefly review the phase space quantization.

Let us consider the one-dimensional system in the state $|\phi\rangle$.
(It is not difficult to generalize to higher dimensional case.)
First we introduce the characteristic operator:
\be
{\hat M}({\hat p},{\hat x};\tau,\tet)=
e^{i(\tau{\hat p}+\tet{\hat x})},
\ee
where $\tau$ and $\tet$ have the same dimension as one of position and
momentum, respectively.
The expectation value $M$ of ${\hat M}$ in state $|\phi\rangle$
can be written as
\be
M(\tau,\tet)=\int dx\ \phi\biggl(x+\bun{\hbar}2\tau\biggr)
\,
\phi^*\biggl(x-\bun{\hbar}2\tau\biggr)\,e^{i\tet x}.
\ee
We define the Wigner distribution as the Fourier transformation of the above
equation:
\be
\fw(p,x)=\bun 1{(2\pi)^2}\int d\tau\int d\tet\ M(\tau,\tet)
\,e^{-i(\tau p+\tet x)}
={1 \over 2\pi}\int dy\ \phi\biggl(x+\bun{\hbar}2y\biggr)\,
\phi^*\biggl(x-\bun{\hbar}2y\biggr)
\,e^{-ipy}.
\lb{wigd}
\ee
The Wigner function corresponds to the classical probability density on
the phase space and reduces to it in the limit of  $\hbar \ra 0$,
but in general it might take negative values and hence does
not have the meaning of a probability density except in the
classical limit.

Now let us consider an arbitrary operator of ${\hat p}$, ${\hat x}$,
say ${\hat A}({\hat p},{\hat x})$ as follows:
\be
{\hat A}({\hat p},{\hat x})
=\int d\tau\int d\tet\ a(\tau,\tet)
\,e^{i(\tau{\hat p}+\tet{\hat x})}.
\ee
In this expression the operator is taken an appropriate
representation, {\it e.g.} $\hat p =-i\hbar\, \pa/\pa\hat x$ in the
position representation.
We assume the corresponding phase space function has the following form
as
\be
A(p,x)
=\int d\tau\int d\tet\ a(\tau,\tet)
\,e^{i(\tau p+\tet x)}.
\ee
This correspondence is called the Weyl correspondence.
Remark that such correspondence fixes the operator ordering
(Weyl ordering).
For example, the classical quantity $p^m x^n$ becomes
\be
\bun 1{2^n}\s_{r=0}^{n}
\left(
\begin{array}{c}
n \\
r
\end{array}
\right)
\,{\hat x}^{n-r}\,{\hat p}^{m}\,{\hat x}^r.
\ee

If we take the expectation value of $\hat A$
in the state  $|\phi\rangle$, we get
\be
\langle \phi| {\hat A} | \phi \rangle =
      \int dp\int dx\ \fw(p,x)\,A(p,x).
\lb{pap}
\ee
The expectation value of the commutator $[\hat A, \hat B]$
becomes
\be
\l\phi|[{\hat A}({\hat p},{\hat x}),
{\hat B}({\hat p},{\hat x})]|\phi\r  =i\hbar
  \int dp\int dx\ \fw(p,x)\,\mbr AB\ .
\ee
The Moyal bracket reduces to the Poisson bracket in
the $\hbar \rightarrow 0$ limit.
In this formalism, the time evolution of the observable is written as
\be
\ddx tA(t)=\mbr {A(t)}H
\lb{moyeq}
\ee
and it corresponds to the Heisenberg picture.
We also consider the Schr\"odinger picture in which
the Wigner distribution depends on time.
It originates from the density matrix $\hat {\rho}$, and
in the statistical treatment the expectation value is denoted by
${\rm Tr} ({\hat \rho}{\hat A})$.
In this sense we can consider the time evolution of the Wigner distribution
and it becomes
\be
\ddx t \fw(t)=\mbr H{\fw(t)}.
\lb{wigeq}
\ee
For detailed treatments of the dynamics of these equations, see
\cite{wig}--\cite{om}.



\vskip 1cm

\noindent{\large\bf Appendix \hspace{-.4in}
\newsection{\hspace{.15in}}}

\vskip -.8cm

In this appendix I show that the relation
$\l \Omega_{\fw}, \xdn{\mbr {A(t)}H}\r=\l \Omega_{\mbr H{\fw(t)}}, \xdn A\r$
holds.
To verify the relation I first prove
\be
\int d\vec x\ \fw(\xdn HA)=-\int d\vec x\ (\xdn H\fw)A.
\lb{b1}
\ee
\noindent {\it Proof of \rf{b1}:}
\bea
&&\hspace*{-1cm}(\mbox{l.h.s. of (B.1)})
=-\int d\vec x\ \fw(\vec x)\mbr AH(\vec x)\nn\\
&=&-\bun 1{(2\pi)^2}\int d\vec x\ \fw(\vec x)\int d\vec a \int d\vec b\
\sin(\vec a\times\vec b)\,A(\vec x+i\vec b)\,H(\vec x+i\vec a)\nn\\
&=&-\bun 1{(2\pi)^2}\int d\vec x\int d\vec a \int d\vec b\
\sin(\vec a\times\vec b)\,\fw(\vec x-i\vec b)\,
H(\vec x+i\vec a-i\vec b)\,A(\vec x)\nn\\
&=&\bun 1{(2\pi)^2}\int d\vec x\int d\vec a \int d\vec b\
\sin(\vec a\times\vec b)\,\fw(\vec x+i\vec b)\,H(\vec x+i\vec a)
\,A(\vec x)\nn\\
&=&\int d\vec x\ \mbr H\fw(\vec x)\,A(\vec x)=(\mbox{r.h.s. of (B.1)})
\hspace{1cm}\Box \nn
\eea

Combining \rf{b1} with the relations as
\bd
A(t)=e^{i\xdn Ht}A, \ \  \fw(t) =e^{-i\xdn Ht}\fw,
\ed
we can make sure the following equality:
\be
\l \Omega_{\fw},\xdn{\mbr{\exp(i\xdn Ht) A}H} \r=
\l \Omega_{\mbr H{\exp(-i\xdn Ht)\fw}},\xdn A\r
\lb{b2}
\ee

\noindent {\it Proof of \rf{b2}:}
\bea
\mbox{(l.h.s. of (B.2))}&=&\int d\vec x\ \fw\mbr{e^{i\xdn Ht}A}H
=\sum_{n=0}^{\infty}{(it)^n\over{n!}}\int d\vec x\ \fw
\mbr{(\xdn H)^n A}H
\nonumber\\
&=&\sum_{n=0}^{\infty}{(it)^n\over{n!}}\int d\vec x\
\fw\Bigl((\xdn H)^{n+1} A\Bigr)\nn\\
&=&-\sum_{n=0}^{\infty}{(it)^n\over{n!}}\int d\vec x\
(\xdn H \fw)\,\Bigl((\xdn H)^{n} A\Bigr)\nn\\
&=&\sum_{n=0}^{\infty}{(it)^n\over{n!}}\int d\vec x\
\Bigl((\xdn H)^2 \fw\Bigr)\,\Bigl((\xdn H)^{n-1} A\Bigr)\nn\\
&=&\cdots\nn\\
&=&(-1)^{n+1}\sum_{n=0}^{\infty}{(it)^n\over{n!}}\int d\vec x\
\Bigl((\xdn H)^{n+1} \fw\Bigr)\,A\nn\\
&=&\int d\vec x\ \mbr H{\fw(t)}\,A=(\mbox{r.h.s. of (B.2)})
\hspace{1cm}\Box \nn
\eea

\newpage

\noindent{\large\bf Appendix \hspace{-.4in}
\newsection{\hspace{.15in}} }

\vskip -.8cm

In this appendix I explain the non-commutative flow of the \kp\
and derive \rf{diso}.

In contrast to the KP flow $\pa_r=\pa/\pa t_r$,
we can consider the flow $\pa_{kl}$  \cite{chen,orl}
which do not commute with each other
but commute with $\pa_r$.
In this sense this additional flow is nothing but the symmetry of the
KP-hierarchy.
First remember that the KP-hierarchy is the following linear
system of equations:
\be
Lw=zw,\hspace{.3in}\pat r w=L^{r}_{+}w\ ,
\lb{lin}
\ee
where the Lax operator $L$ and the Baker-Akhiezer function $w$ are
respectively given by
\bea
&&L(x,t)=\pa+\s_{j=1}^{\infty} u_j(x,t)\pa^{-j}, \hspace{1cm}(\pa=\pa/\pa x)\\
&&w(z,t)=\biggl(1+\s_{j=1}^{\infty}w_j \pa^{-j}\biggr)\,
{\rm exp}\s_{r}t_{r}z^{r}=: We^{\xi},
\lb{ba}
\eea
and $(\ )_+$ denotes the purely differential operator part.
We also describe $L$ by using the dressing operator $W$ as $W\pa W$.
The first equation of \rf{lin} means that the operation of $L$ on the
Baker-Akhiezer function is equivalent to the
production of $z$ to $w$.
Similarly, the derivation $\pa_{z}$ to $w$ is
written in terms of the pseudo-differential operators as
\be
\pxpy w z =W\biggl(\ \s_{r=1}^{\infty}rt_{r}\pa^{r-1}\biggr)
W^{-1}We^{\xi}=: Mw\ .
\ee
Then we get $z^{k}\pa_{z}^{l}\ w=M^{l}L^{k} w\ (k\in{\bf Z},\ l\in{\bf Z}_{\geq
0})$.
It enables us to consider the vector field as
\be
\pat{k\,l}\,w=-(M^{l}L^{k})_{-}w.
\lb{new2}
\ee
Moreover between $z^{k}\pa_{z}^{l}$ and $\pat{k\,l}$,
there is a Lie algebra isomorphism $z^{k}\pa_{z}^{l}\mapsto\pat{k\,l}$.
Therefore $\pat{k\,l}$ yields the relation as
\be
[\pat{k\,l},\pat{k'\,l'}]
=\s_{j=1}^{\infty}\Biggl\{
\comb kj \comb lj - \comb{k'}j\comb{l'}j
\Biggr\}\ j!\ \pat{k+k'-j,\ l+l'-j}\ .
\ee
In general the operators spanned by the differential operators
$\{ z^{k}\pa_{z}^{l}; k\in{\bf Z}, l\in{\bf Z}_{\geq 0}\}$
form an infinite-dimensional Lie algebra, and it is called \w \cite{pop}.

Since $w\ (w^*)$ is rewritten by use of the vertex operator $V\ (V^*)$
and the \ta function as
\bea
&&w(z,t)=\bun{V(z)\ \tau}{ \tau}=\bun{\tau (t-1/[z])}{\tau (t)}\ e^{\xi},
\ \ \
w^*(z,t)=\bun{V^*(z)\ \tau}{ \tau }
=\bun{\tau (t+1/[z])}
{\tau (t)} \ e^{-\xi},\nn\\
&&(t\pm 1/[z]:=t_{1}\pm 1/z,\,t_{2}\pm 1/2z^{2},\,\cdots)\nn
\eea
we obtain the action of $\pat{k\,l}$ on the \ta function as
\be
\pa_{m+l,\,l}\,\pa\ln\tau=
\oint \bun{dz}{2\pi i}\ \biggl(z^{m+l}\pa_z^{l}w(z)\biggr)\,
w^{*}(z)
=
\left[\oint \bun{dz}{2\pi i}\ \biggl(z^{m+l}\pa_z^{l}V(z)\biggr)\,
V^*(z)\right]\,\ppx x\ln\tau
\lb{cn}
\ee
\cite{jpa,ow,dic2}.

The derivation of \rf{diso} is shown as follows:
\bea
&&{\rm (r.h.s.\ of\ \rf{zfou})}
=
-\int dx
\oint \bun{dz}{2\pi iz}\ \biggl(z^{m} q^{n(\zp+\bun {m} 2)}w(z)\biggr)\,
w^{*}(z)\nn\\
&&\hspace{1cm}=
-q^{\bun{nm}2}\s_{j,l=0}^{\infty}c_{jl}\bun{(n\lam)^j}{j!}\,
\int dx\oint \bun{dz}{2\pi iz}\ \biggl(z^{m+l}\pa_z^{l}w(z)\biggr)\,
w^{*}(z)\nn\\
&&\hspace{1cm}=
\int dx\ \biggl[q^{\bun{nm}2}\s_{j,l=0}^{\infty}c_{jl}\bun{(n\lam)^j}{j!}\,
\pa_{m+l-1,l}\biggr]\biggl(-\ppx x\ln\tau\biggr)\nn\\
&&\hspace{1cm}=
\int dx\ {\cal D}_{mn}\,\biggl(-\ppx x\ln\tau\biggr)
\hspace{.5cm}{\rm \rf{diso}}\ \ \Box\nn
\lb{expand}
\eea

\vskip 1.7cm

\renewcommand{\Large}{\large} 
\end{document}